\documentclass[reprint,noshowpacs,noshowkeys,prd,balancelastpage,nofootinbib]{revtex4}
\usepackage{amsfonts}
\usepackage{mdframed}
\usepackage{amssymb}
\usepackage{footnote}
\usepackage{amsmath}
\usepackage{graphicx}
\usepackage{float}
\usepackage[font={footnotesize,it}]{caption}
\usepackage[utf8]{inputenc}
\usepackage{natbib}
\usepackage{tikz}
\usepackage{tikz-3dplot}
\usepackage[colorlinks=true,
            linkcolor=purple,
            urlcolor=purple,
            citecolor=blue]{hyperref}

\begin{document}

\title{Causal structure and the geodesics in the hairy extension of the
Bertotti-Robinson spacetime}
\author{Vahideh Memari }
\email{vahideh.memari@emu.edu.tr}
\author{S. Habib Mazharimousavi}
\email{habib.mazhari@emu.edu.tr, Corresponding author}
\affiliation{Department of Physics, Faculty of Arts and Sciences, Eastern Mediterranean
University, Famagusta, North Cyprus via Mersin 10, T\"{u}rkiye}
\date{\today }

\begin{abstract}
A hairy extension of the Bertotti-Robinson regular spacetime has been
recently introduced in the context of the Einstein-Maxwell-Scaler theory
that surprisingly is a singular black hole formed in the $S_{3}$ background
spatial topology [CQG39(2022)167001]. In this research, we first clarify the
topology of the spacetime based on the coordinate transformations as well as
the energy-momentum configuration and the causal structure of the black
hole. Furthermore, we investigate the geodesics of the null and timelike
particles in this spacetime. It is shown that in the radial motion on the
equatorial plane, while photons may collapse to the singularity or escape to
the edge of the universe, a massive particle always collapses to the
singularity. The general geodesics of null and massive particles reveal that
all particles except the outgoing light ray, eventually fall into the black
hole.
\end{abstract}

\keywords{Black hole; Geodesics; Causal structure;}
\maketitle

\section{Introduction}

In general relativity, no-hair theorems state that black holes are described
by only three parameters, their mass $M$, their electromagnetic charge $Q$,
and their angular momentum $\ell $ \cite{1,2,3,4,5}. Recent studies on black
holes indicate the existence of hairy black holes some of which form in the
presence of Yang-Mills fields \cite{7,8,9}. Hairy black holes are solutions
to Einstein's equations in the interaction of diverse kinds of matter fields
with gravity \cite{10}. For instance gravity in interaction with
electromagnetism and axion \cite{11,12,13}, gravity coupled to
electromagnetism and dilaton \cite{14,15,16,17}, and gravity in interaction
with electromagnetism and Abelian Higgs field \cite{18}. The first study on
the coupling of gravity and a scalar field was considered by Fisher who
presented a static and spherically symmetric solution in the theory \cite%
{Fisher}. It is known that the string theory in the low energy limit reduces
to Einstein's theory coupled nontrivially to other fields such as axion,
dilaton, and gauge fields \cite{Dil}. Such coupling for instance in
Einstein-Maxwell-Dilaton (EMD) theory changes significantly the physical
characteristics of the black hole such as the causal structure, the
asymptotic behavior and the thermodynamics from the Reissner-Nordstr\"{o}m
black hole \cite{Dil2,Dil3}. Several exact black hole solutions in EMD
theory have been obtained that are either asymptotically flat \cite%
{AF1,AF2,AF3,AF4,AF5,AF6,AF7} or non-asymptotically flat \cite%
{NF1,NF2,NF3,NF4,NF5,NF6,NF7,NF8,NF9,NF10,NF11,NF12,NF13,NF14,NF15}. Among
the non-asymptotically flat solutions, some are asymptotically de-Sitter or
anti-de Sitter \cite{ds1,ds2,ds3,ds4,ds5,ds6,ds7,ds8,ds9,ds10,ds11} which
have applications in the anti-de Sitter, conformal field theory
correspondence (AdS/CFT) in the context of the unification of the quantum
field theory and gravitation \cite{CFT}. We note that EMD is not the only
extension of the Einstein-Maxwell (EM) theory and the coupling of gravity to
scalar fields through Maxwell's gauge i.e., Einstein-Maxwell-Scalar (EMS)
theory, has also received great attention in the literature \cite%
{EMS,S2,S3,S4,S5,S6,S7,S8,S9,S10}. For a good classification of EMS theory,
we refer to a recent paper by Astefanesei et al \cite{S11}. Besides the
theory of scalar/dilaton coupled to gravity, the scalar-tensor theory of
gravity with the Higgs field as the scalar field has been studied in
different contexts such as the flat rotation curves anomaly of spiral
galaxies \cite{H1}, and the Higgs field as quintessence \cite{H2}. In \cite%
{H3} horizon-less spherically symmetric vacuum solutions have been
introduced in the Higgs scalar-tensor theory of gravity such that in the
vanishing limit of the Higgs field excitations, the usual Schwarzschild
black hole is recovered.

In this current study, we investigate further the hairy extension of the
Bertotti-Robinson (BR) spacetime \cite{19,20,21,22,23,24,R1,R2} in the
context of the EM theory \cite{25}, a subgroup of hairy black holes which
have gravity in interaction with electromagnetism and scalar fields \cite{26}%
. This new spacetime eventuates a black hole in closed space. One of the
reasons to study a closed space in a cosmological sense is that the energy
and the topology of an open universe are difficult to be determined.
Moreover, a static, closed space can depict information related to the
dynamics of the universe and an early universe \cite{27}. One of the
well-known cosmological models of the universe is the Robertson-Walker (RW) 
\cite{WR} metric which is described by%
\begin{equation}
ds_{RW}^{2}=-dt^{2}+a\left( t\right) ^{2}\left( dr^{2}+S_{K}\left( r\right)
^{2}\left( d\theta ^{2}+\sin ^{2}\theta d\varphi ^{2}\right) \right)
\label{U1}
\end{equation}%
where 
\begin{equation}
S_{K}\left( r\right) =\left\{ 
\begin{array}{ccc}
r &  & K=0 \\ 
\sin r &  & K=+1 \\ 
\sinh r &  & K=-1%
\end{array}%
\right.  \label{U2}
\end{equation}%
and $K$ stands for the spatial Gaussian curvature. Considering the topology
of the universe (\ref{U1}) to be $%
\mathbb{R}
\times \Sigma $ with $%
\mathbb{R}
$ representing the time and $\Sigma $ the spatial three-manifold, $K=0,-1$
and $+1$ imply that $\Sigma $ is flat, open or closed, respectively.
Applying the transformation given by $S_{K}\left( r\right) ^{2}\rightarrow
R^{2}$ the line element (\ref{U1}) becomes%
\begin{equation}
ds_{RW}^{2}=-dt^{2}+a\left( t\right) ^{2}\left( \frac{dR^{2}}{1-KR^{2}}%
+R^{2}\left( d\theta ^{2}+\sin ^{2}\theta d\varphi ^{2}\right) \right) .
\label{U3}
\end{equation}%
With $K=+1,$ (\ref{U3}) represents a closed universe in the sense that $R$
is confined to $R\in \left[ 0,\sqrt{\frac{1}{K}}\right) .$ Furthermore,
assuming $a\left( t\right) =a_{0}=const.,$ (\ref{U3}) becomes a static
closed universe upon which Einstein's field equation $G_{\mu }^{\nu }=T_{\mu
}^{\nu }$ yields%
\begin{equation}
T_{\mu }^{\nu }=diag\left[ -\rho _{0},p_{0},p_{0},p_{0}\right]  \label{U4}
\end{equation}%
where the energy density and the pressure are given by $\rho _{0}=\frac{3K}{%
a_{0}^{2}}$ and $p_{0}=-\frac{K}{a_{0}^{2}},$ respectively. The equation of
state (EoS) of the perfect fluid supporting the above static universe is
given by 
\begin{equation}
p_{0}=-\frac{\rho _{0}}{3}.  \label{U5}
\end{equation}%
It is clear that with $K=0$ the flat universe is\ a vacuum, with $K=+1$ the
closed universe is supported by a physical isotropic matter field whose
positive energy density (with finite total energy) and negative isotropic
pressure (\textbf{with vanishing gravitational mass density,} $\rho
_{G}=\rho _{0}+3p_{0}$) satisfy all the energy conditions including the null
energy condition ($\rho _{0}+p_{0}\geq 0$), the weak energy condition ($\rho
_{0}\geq 0,\rho _{0}+p_{0}\geq 0$) and the strong energy condition ($\rho
_{0}+3p_{0}\geq 0$). On the other hand, with $K=-1$ the open universe is
supported by an exotic matter with a negative energy density (with an
infinite total exotic matter) and a positive isotropic pressure (which
pushes the universe away from the center), and all energy conditions are
violated.

Studying the geodesics of the massless or massive particles in the vicinity
of the black holes plays a great role in understanding the structure of
their spacetime geometry. Such investigations around different black holes
have been published in literature ever since the first black hole was
introduced. Here \cite{G1,G2,G3,G4,G5,G6,G7,G8,G9,G10,G11,G12,G13,G14,G15},
we refer to some of them and the interested reader can find more in their
references as well. In this context, we are going to study geodesics
equations of a test particle in the spacetime of the black hole powered by
pure magnetic fields introduced in \cite{25}. The main streamline of the
present work is to investigate the trajectory of massless/null and
massive/timelike test particles. We derive the energy and angular momentum
of the test particles and we illustrate some observable quantities in graphs
to discover the particle's behavior in different circumstances.

The organization of the paper is as follows. In Sec. II we present a review
of the action, the field equations, and the black hole solution obtained
previously in \cite{25}. Sec. III is devoted to the topology of the
mentioned spacetime and in particular, we concentrate on the formation of
the black hole in $S_{3}$ background spatial topology, from different
aspects including the energy conditions, the causal structure, and the
Penrose diagram. In Sec. IV the geodesics equation of the null and massive
particles are analyzed and the results are presented either analytically or
numerically through some graphs. Finally in Sec. V the paper is concluded by
remarking on the results.

\section{A review on the hairy extension of the BR spacetime}

The black hole solution reported in \cite{25} has been obtained in the
context of EMS theory where the action is given by ($8\pi G=c=1$)%
\begin{equation}
I=\frac{1}{2}\int d^{4}x\sqrt{-g}\left( \mathcal{R}-2\partial _{\mu }\psi
\partial ^{\mu }\psi +V_{0}\cos \left( \frac{\psi }{\sqrt{2}}\right) F_{\mu
\nu }F^{\mu \nu }\right)  \label{I1}
\end{equation}%
in which $\psi $ is the scalar field, $V_{0}$ is a positive coupling
constant, $\mathcal{R}$ is the Ricci scalar and $F_{\mu \nu }F^{\mu \nu }$
is Maxwell's invariant. Variation of the action with respect to $g_{\mu \nu
},$ $A_{\mu }$ and $\psi ,$ respectively, yields the field equations given
by 
\begin{equation}
\mathcal{R}_{\mu }^{\nu }=2\partial _{\mu }\psi \partial ^{\nu }\psi +\frac{1%
}{2}V_{0}\cos \left( \frac{\psi }{\sqrt{2}}\right) \left( 4F_{\mu \lambda
}F^{\nu \lambda }-F_{\alpha \beta }F^{\alpha \beta }\delta _{\mu }^{\nu
}\right) ,
\end{equation}%
\begin{equation}
\nabla _{\mu }\left( V_{0}\cos \left( \frac{\psi }{\sqrt{2}}\right) F^{\mu
\nu }\right) =0
\end{equation}%
and%
\begin{equation}
\nabla _{\mu }\nabla ^{\mu }\psi =-\frac{V_{0}}{4\sqrt{2}}\sin \left( \frac{%
\psi }{\sqrt{2}}\right) F_{\mu \nu }F^{\mu \nu }.
\end{equation}%
The electromagnetic gauge field is considered to be a pure radial magnetic
field produced by a magnetic monopole sitting at the origin such that%
\begin{equation}
\mathbf{F}=P\sin \theta d\theta \wedge d\varphi
\end{equation}%
where the constant parameter $P$ is the magnetic charge. For the method upon
which the field equations have been solved, we refer to the original paper
in Ref. \cite{25} where all the details can be found. Hence, we only write
the solutions which are as follow. The spacetime is spherically symmetric
with the line element 
\begin{equation}
ds^{2}=-f\left( r\right) dt^{2}+\frac{dr^{2}}{f\left( r\right) }+R\left(
r\right) {}^{2}d\Omega ^{2},  \label{I2}
\end{equation}%
in which the metric functions are given by

\begin{equation}
f\left( r\right) =\frac{\left( r+r_{h}\right) {}^{3}\left( r-r_{h}\right) }{%
R_{0}^{2}r^{2}},  \label{I3}
\end{equation}%
and

\begin{equation}
R\left( r\right) =R_{0}\left( \frac{r}{r+r_{h}}\right) .  \label{I4}
\end{equation}%
Furthermore, the radial magnetic field and the scalar field are obtained to
be%
\begin{equation}
B\left( r\right) =\frac{P\left( r+r_{h}\right) ^{2}}{R_{0}^{2}r^{2}},
\label{I5}
\end{equation}%
and 
\begin{equation}
\psi \left( r\right) =\pm 2\sqrt{2}\arctan \left( \sqrt{\frac{r}{r_{h}}}%
\right) ,  \label{I6}
\end{equation}%
respectively. Herein, $r_{h}$ is the event horizon and $R_{0}^{2}=P^{2}V_{0}$%
. The scalar field (\ref{I6}) becomes constant as $r_{h}\rightarrow 0$ i.e., 
\begin{equation}
\psi \left( r\right) \rightarrow \pm \sqrt{2}\pi \text{ \ as }%
r_{h}\rightarrow 0  \label{I7}
\end{equation}%
and consequently 
\begin{equation}
V_{0}\cos \left( \frac{\psi }{\sqrt{2}}\right) F_{\mu \nu }F^{\mu \nu
}\rightarrow -V_{0}F_{\mu \nu }F^{\mu \nu }  \label{I8}
\end{equation}%
which upon setting $V_{0}=1$ the action reduces to EM theory, and the
spacetime becomes%
\begin{equation}
ds^{2}=-\frac{r^{2}}{P^{2}}dt^{2}+\frac{P^{2}dr^{2}}{r^{2}}+P^{2}d\Omega
^{2}.  \label{I9}
\end{equation}%
This line element is not the Reissner-Nordstr\"{o}m black hole solution of
the EM theory. Instead, the transformations $r=\frac{1}{\tilde{r}}$ and $t=P%
\tilde{t}$ yield%
\begin{equation}
ds^{2}=\frac{P^{2}}{\tilde{r}^{2}}\left( -d\tilde{t}^{2}+d\tilde{r}^{2}+%
\tilde{r}^{2}d\Omega ^{2}\right)  \label{I10}
\end{equation}%
which reveals that (\ref{I9}) is the regular solution of the EM theory known
as the BR spacetime \cite{19,20}.

\begin{figure}[tbp]
\includegraphics[scale=0.7]{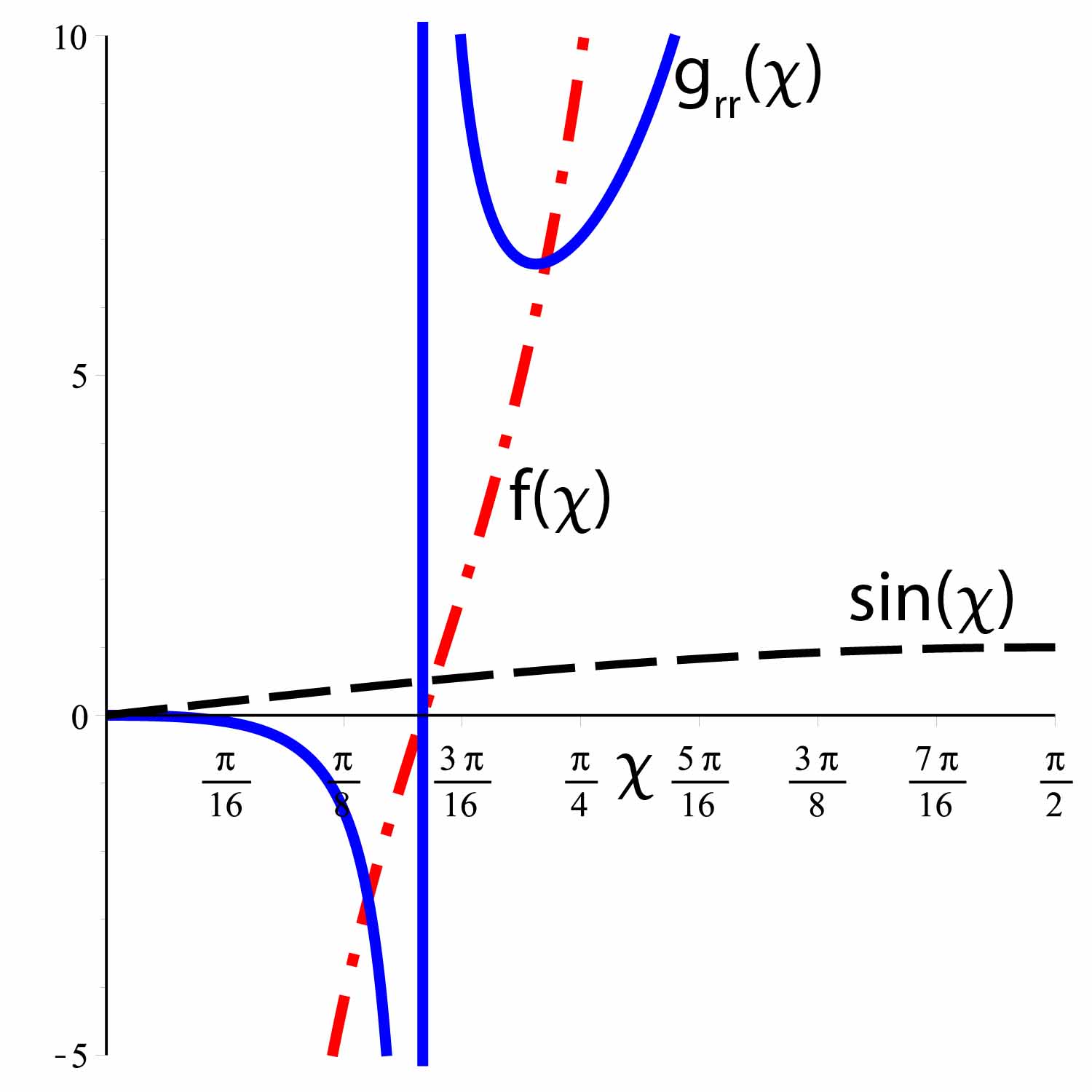}
\caption{The plots of scaled transformed metric function i.e., $\frac{%
R_{0}^{2}}{r_{h}^{2}}f\left( \protect\chi \right) $ (red, dash-dot)$,$ as
well as $\frac{1}{R_{0}^{2}}g_{\protect\chi \protect\chi }$ from (\protect
\ref{K7}) (blue, solid) and $\sin \protect\chi $ (dash) for $0\leq \protect%
\chi \leq \frac{\protect\pi }{2}$.}
\label{F0}
\end{figure}

\section{Black hole in the $S_{3}$ space topology background}

\begin{figure}[tbp]
\includegraphics[scale=0.7]{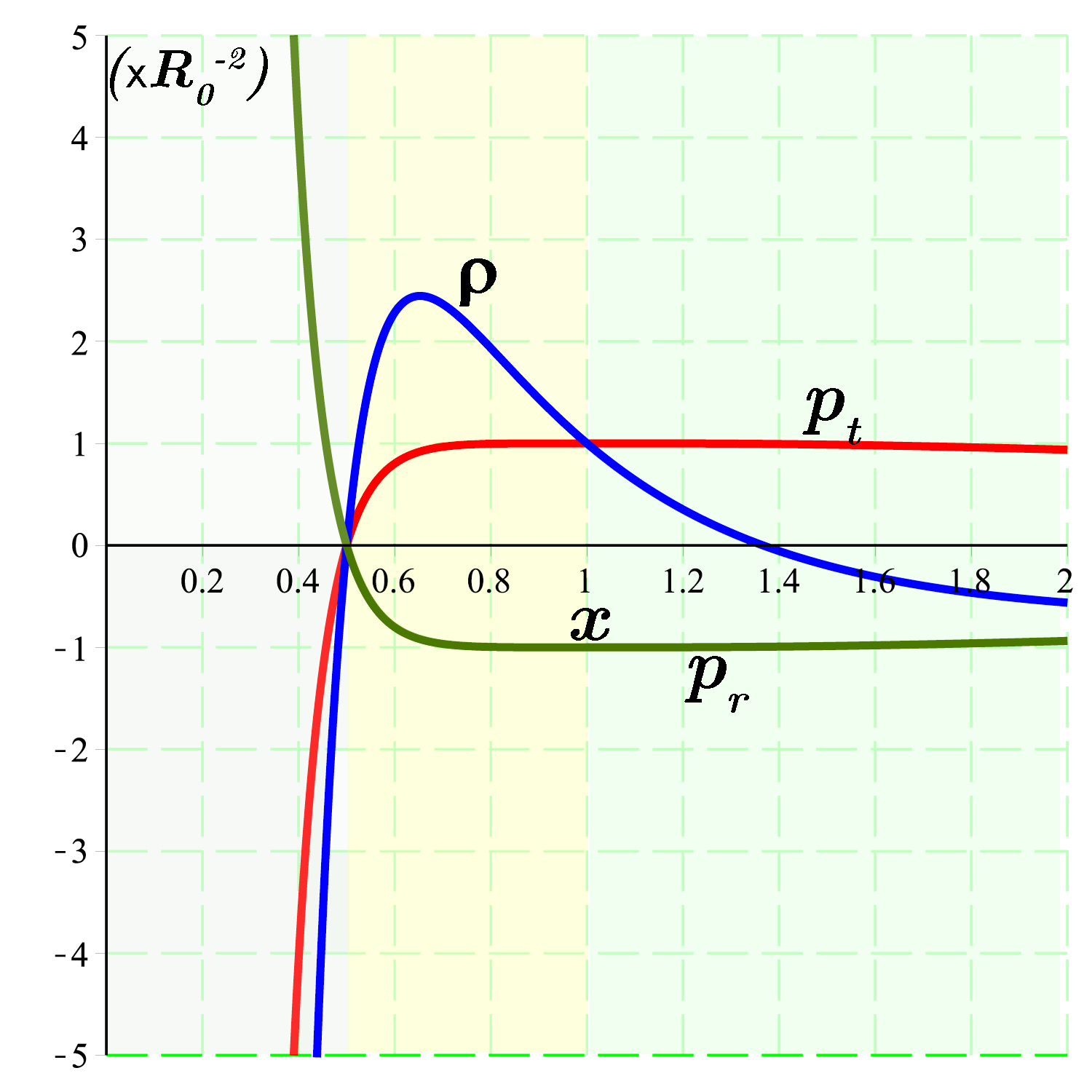}
\caption{Plots of scaled $\protect\rho ,p_{r}$ and $p_{t}$ in terms of $x=%
\frac{R}{R_{0}}.$ The negative radial pressure outside the horizon is in
analogy with the RW closed universe.}
\label{F1}
\end{figure}
\begin{figure}[tbp]
\includegraphics[scale=0.7]{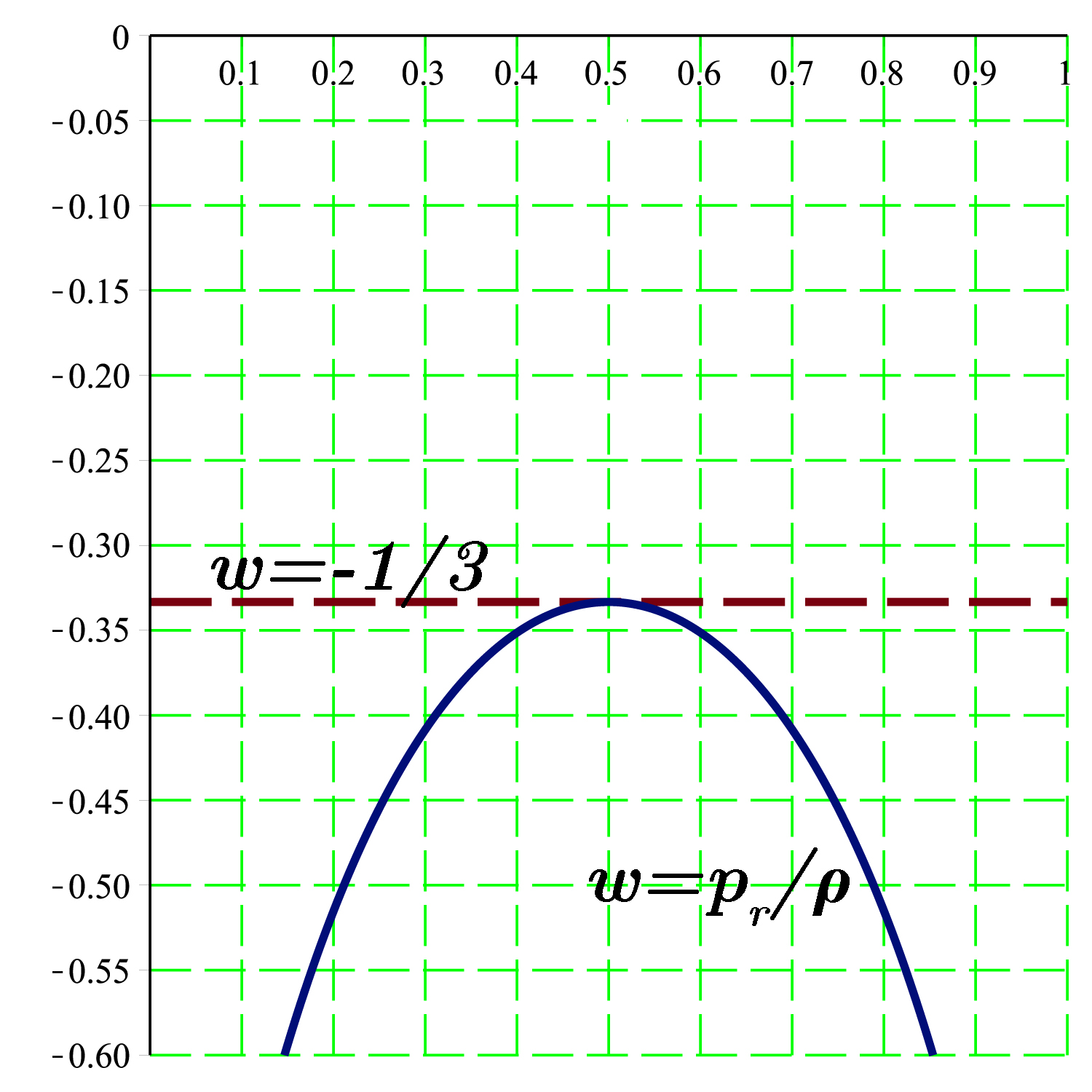}
\caption{Plot $w=\frac{p_{r}}{\protect\rho }$ in terms of $x=\frac{R}{R_{0}}%
. $ Near the horizon, it approaches $-\frac{1}{3}$ which is the same as the
static RW closed universe.}
\label{F2}
\end{figure}
\begin{figure}[tbp]
\includegraphics[scale=0.7]{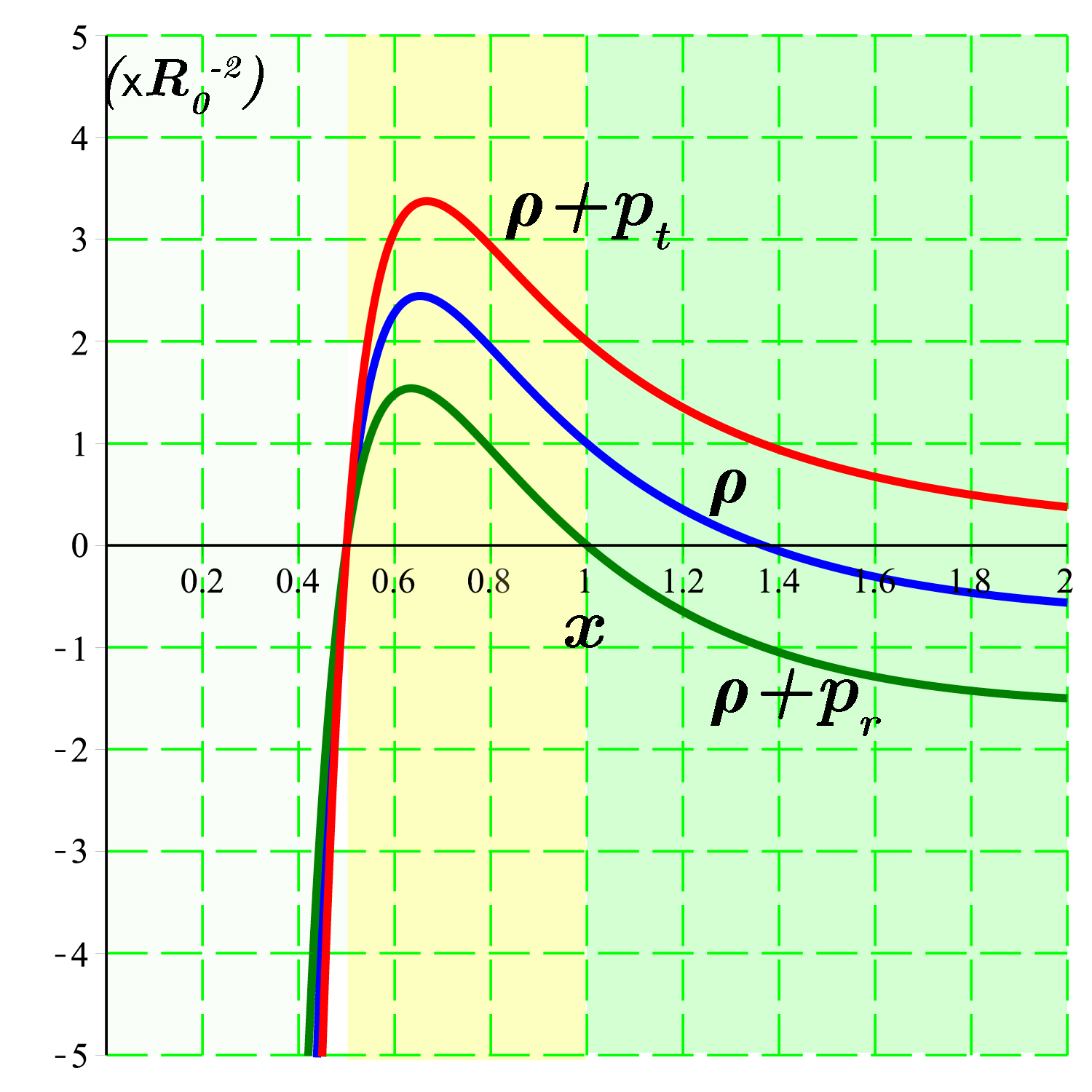}
\caption{Plots of scaled $\protect\rho ,\protect\rho +p_{r}$ and $\protect%
\rho +p_{t}$ in terms of $x=\frac{R}{R_{0}}.$ While all of them are positive
outside the horizon within the boundary of the universe at $x=1$, beyond
this boundary we observe $\protect\rho +p_{r}<0$, which implies that all
energy conditions are violated and therefore the fluid is exotic. }
\label{F3}
\end{figure}

In \cite{25} where the black hole solution (\ref{I2}) has been introduced,
it was claimed that this black hole is closed in 3-space or equivalently, it
is formed in the $S_{3}$ background spatial topology. Such a black hole is
known in the literature (see for instance \cite{28,29,30,31,32,33}). To
understand this feature we would like to make some investigation through the
black holes which have been considered in this category (in particular, we
refer to section 3.1 of Ref. \cite{31}). To do so we again start with the
static cosmological model expressed by (\ref{U3}) which is actually a pure $%
S_{3}$ space topology solution. With $a\left( t\right) =1$ the spatial part
becomes%
\begin{equation}
ds_{3}^{2}=\frac{dR^{2}}{1-KR^{2}}+R^{2}\left( d\theta ^{2}+\sin ^{2}\theta
d\varphi ^{2}\right)  \label{K1}
\end{equation}%
which after introducing $R=\frac{R_{0}}{\sqrt{K}}\sin \chi $ it becomes%
\begin{equation}
ds_{3}^{2}=\frac{1}{K}R_{0}^{2}\left( d\chi ^{2}+\sin ^{2}\chi \left(
d\theta ^{2}+\sin ^{2}\theta d\varphi ^{2}\right) \right)  \label{K2}
\end{equation}%
with the trivial topology of $S_{3}.$ We note that (\ref{K2}) possesses the $%
S_{3}$ space topology for all values of $\chi \in \left[ 0,\pi \right] $ and
that is the reason it is pure $S_{3}$ with the geometry of 3-sphere.

In addition to the static RW spacetime which is the solution of Einstein's
equation supported by an isotropic, homogeneous, and uniform perfect fluid
with the energy-momentum tensor governed by the EoS given by (\ref{U5}),
there are black hole solutions \cite{28,29,30,31,32,33} which are supported
by non-uniform perfect fluid with an EoS given by%
\begin{equation}
p\left( r\right) =-\frac{\rho \left( r\right) }{3}.  \label{K3}
\end{equation}%
One such solution was introduced by Bronnikov and Zaslavskii in \cite{28}
and can be described by the line element%
\begin{equation}
ds^{2}=-\left( 1-k\cot \chi \right) dt^{2}+\frac{R_{0}^{2}}{\left( 1-k\cot
\chi \right) }d\chi ^{2}+R_{0}^{2}\sin ^{2}\chi \left( d\theta ^{2}+\sin
^{2}\theta d\varphi ^{2}\right)  \label{K4}
\end{equation}%
in which $k$ and $R_{0}$ are two constants and $0\leq \chi \leq \pi .$ It is
a black hole solution powered by the non-uniform perfect fluid with the
energy-momentum tensor given by%
\begin{equation}
T_{\mu }^{\nu }=\frac{3k\cot \chi -\sin \chi }{R_{0}^{2}\sin \chi }%
diag\left( 1,-\frac{1}{3},-\frac{1}{3},-\frac{1}{3}\right) .
\end{equation}%
The energy-momentum tensor satisfies the EoS (\ref{K3}), whereas the
topology of the black hole (\ref{K4}) is not pure $S_{3}.$ However, in the
vicinity of $\chi =\frac{\pi }{2},$ the spatial part of the spacetime becomes%
\begin{equation}
ds_{3}^{2}\simeq R_{0}^{2}\left( d\chi ^{2}+\sin ^{2}\chi \left( d\theta
^{2}+\sin ^{2}\theta d\varphi ^{2}\right) \right)  \label{K5}
\end{equation}%
which reveals the $S_{3}$ space topology. We note that the geometry of the
spatial part of (\ref{K4}) away from $\chi =\frac{\pi }{2}$ is nontrivial.
In such a configuration, the spacetime is interpreted to be "\textit{formed
in the }$S_{3}$\textit{\ background spatial topology}" (see \cite{31} for
more details).

Regarding the black hole solution subjected in this study i.e., the line
element (\ref{I3}), by introducing $\frac{r}{r+r_{h}}=\sin \chi $ the
spatial segment transforms into%
\begin{equation}
ds_{3}^{2}=\frac{R_{0}^{2}\sin ^{2}\left( 2\chi \right) d\chi ^{2}}{4\left(
1-\sin \chi \right) ^{2}\left( 2\sin \chi -1\right) }+R_{0}{}^{2}\sin
^{2}\chi d\Omega ^{2}  \label{K6}
\end{equation}%
in which $0\leq \chi \leq \frac{\pi }{2}$ and the horizon is located at $%
\chi =\frac{\pi }{6}.$ With this setting the geometry of the spacetime is
still nontrivial although in the vicinity of $\chi \simeq \frac{14835\pi }{%
65536},$ (\ref{K6}) approximately becomes%
\begin{equation}
ds_{3}^{2}\simeq R_{0}^{2}\left( \left( 1+\zeta ^{2}\right) d\chi ^{2}+\sin
^{2}\chi d\Omega ^{2}\right)  \label{K7}
\end{equation}%
which reveals the $S_{3}$ space topology of the spacetime near $\chi \simeq 
\frac{14835\pi }{65536}$ with a correction $\zeta ^{2}\simeq 5.6379.$ In
Fig. \ref{F0}, we plot the scaled transformed metric function i.e., $\frac{%
R_{0}^{2}}{r_{h}^{2}}f\left( \chi \right) ,$ as well as $\frac{1}{R_{0}^{2}}%
g_{\chi \chi }$ from (\ref{K7}) and $\sin \chi $ for $0\leq \chi \leq \frac{%
\pi }{2}.$ It is observed that $\frac{1}{R_{0}^{2}}g_{\chi \chi }$ around
its minimum i.e., $\chi \simeq \frac{14835\pi }{65536}$ is almost constant.

To get a better idea of the topology of the black hole spacetime (\ref{I2})
let us transform the spacetime into a more familiar system of coordinates.
To do so, we introduce the following transformation 
\begin{equation}
R\left( r\right) =R_{0}\left( \frac{r}{r+r_{h}}\right) =R  \label{r1}
\end{equation}%
upon which the line element (\ref{I2}) becomes%
\begin{equation}
ds^{2}=-f\left( R\right) dt^{2}+\frac{dR^{2}}{h\left( R\right) }%
+R^{2}d\Omega ^{2}  \label{r2}
\end{equation}%
in which%
\begin{equation}
f\left( R\right) =\frac{2r_{h}R_{0}\left( R-R_{0}/2\right) }{R^{2}\left(
R_{0}-R\right) ^{2}}  \label{r3}
\end{equation}%
and%
\begin{equation}
h\left( R\right) =\frac{2\left( R_{0}-R\right) ^{2}\left( R-R_{0}/2\right) }{%
R_{0}R^{2}}.  \label{r4}
\end{equation}%
We note that, since $r\in \left[ 0,\infty \right) ,$ (\ref{r1}) implies $%
R\in \left[ 0,R_{0}\right) $ and indeed $t\in \left( -\infty ,\infty \right) 
$ remains unchanged. Clearly (\ref{r2}) is a black hole with an event
horizon located at $R=R_{+}=\frac{R_{0}}{2}.$ One can show that the Ricci
and the Kretschmann scalars are respectively given by%
\begin{equation}
\mathcal{R}=\frac{4\left( 2R-R_{0}\right) \left( R_{0}-R\right) }{RR^{3}}
\label{r5}
\end{equation}%
and%
\begin{equation}
\mathcal{K}=\frac{8\left(
30R^{6}-120R^{5}R_{0}+214R^{4}R_{0}^{2}-216R^{3}R_{0}^{3}+132R^{2}R_{0}^{4}-46RR_{0}^{5}+7R_{0}^{6}\right) 
}{R^{8}R_{0}^{2}}  \label{r6}
\end{equation}%
which imply that the black hole is singular at $R=0$ and the Ricci scalar is
positive outside the horizon. Let us add also that although the signature of
the spacetime doesn't change even with $R>R_{0}$ which indicates (\ref{r2})
is physically valid even for $R>R_{0}$, the scalar field $\psi $ transforms
into%
\begin{equation}
\psi \left( R\right) =\pm 2\sqrt{2}\arctan \left( \sqrt{\frac{R}{R_{0}-R}}%
\right)
\end{equation}%
which obviously is a pure imaginary function. This in turn indicates that
the nature of the scalar field $\psi \left( R\right) $ changes into a
phantom field. Similar to the RW universe (\ref{U1}), we apply Einstein's
equation for the black hole metric i.e. (\ref{r2}) to obtain%
\begin{equation}
T_{\mu }^{\nu }=G_{\mu }^{\nu }=\left( -\rho ,p_{r},p_{t},p_{t}\right)
\end{equation}%
in which the energy density $\rho ,$ the radial pressure $p_{r},$ and the
transverse pressure $p_{t}$ are, respectively, given by%
\begin{equation}
\rho =\frac{6x^{2}-4x^{3}-1}{R_{0}^{2}x^{4}},
\end{equation}%
\begin{equation}
p_{r}=\frac{6x^{2}-4x^{3}-4x+1}{R_{0}^{2}x^{4}}
\end{equation}%
and%
\begin{equation}
p_{t}=\frac{\left( 2x-1\right) \left( 2x^{2}-2x+1\right) }{R_{0}^{2}x^{4}}
\end{equation}%
in which $x=\frac{R}{R_{0}}.$ In Fig. \ref{F1}, we plot $\rho $, $p_{r}$ and 
$p_{t}$ (all scaled by $R_{0}^{2}$) in terms of $x.$ It is observed that at
the horizon they all vanish and near the horizon, the EoS approaches $p_{r}=-%
\frac{1}{3}\rho .$ This can be seen in Fig. \ref{F2} where we plot $w=\frac{%
p_{r}}{\rho }$ in terms of $x$ near the horizon. A similar EoS was already
observed in the static RW universe in Eq. (\ref{U5}). Moreover, the radial
pressure is negative outside the horizon which is in analogy with the RW
closed universe. In Fig. \ref{F3}, we plot also $\rho $, $\rho +p_{r}$ and $%
\rho +p_{t}$ (all are scaled by $R_{0}^{2}$) in terms of $x$. It is observed
that since $\rho +p_{r}$ is negative for $x>1$ (outside the boundary of the
universe) non of the energy conditions are satisfied. This implicitly
implies that the physical region is located only within $x<1.$

\subsection{Causal structure}

\begin{figure}[tbp]
\includegraphics[scale=0.7]{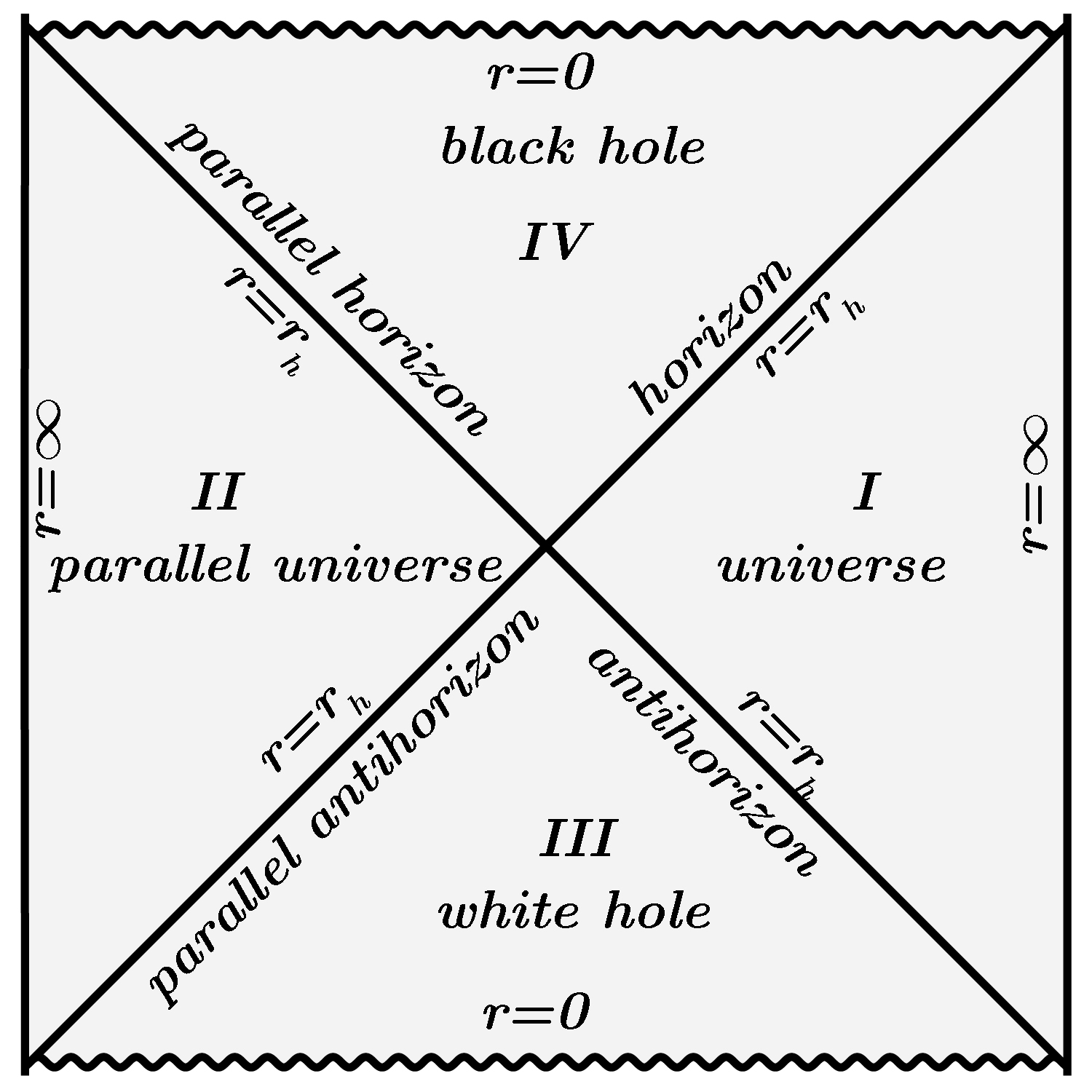}
\caption{ Penrose diagram of the black hole spacetime, given in (\protect\ref%
{I2}). The spacetime singularity is spacelike and is located at $r=0$.
Different regions are labeled appropriately on the figure.}
\label{F4}
\end{figure}

To obtain the causal structure of the spacetime, we start from the original
line element given in (\ref{I2}) and obtain the tortoise coordinate ($%
r>r_{h} $)%
\begin{equation}
r_{\ast }=\int \sqrt{-\frac{g_{rr}}{g_{tt}}}dr=\int \frac{dr}{f\left(
r\right) }=-\frac{R_{0}^{2}}{8r_{h}}\left[ \frac{2r_{h}\left(
3r+2r_{h}\right) }{\left( r+r_{h}\right) ^{2}}+\ln \left\vert \frac{r+r_{h}}{%
r-r_{h}}\right\vert \right] ,  \label{r7}
\end{equation}%
in which while $r\in \left( r_{h},\infty \right) $, $r_{\ast }\in \left(
-\infty ,0\right) .$ Introducing the null coordinates i.e., $u=t-r_{\ast }$
and $v=t+r_{\ast }$ yields%
\begin{equation}
ds^{2}=-\frac{\left( r+r_{h}\right) {}^{3}\left( r-r_{h}\right) }{%
R_{0}^{2}r^{2}}dudv+R\left( r\right) {}^{2}d\Omega ^{2}  \label{r8}
\end{equation}%
where $r=r\left( u,v\right) .$ Now, we employ the Kruskal-Szekeres
coordinates defined by 
\begin{equation}
U=-\frac{R_{0}^{2}}{4r_{h}}e^{-\frac{4r_{h}}{R_{0}^{2}}u}  \label{r9}
\end{equation}%
and%
\begin{equation}
V=\frac{R_{0}^{2}}{4r_{h}}e^{\frac{4r_{h}}{R_{0}^{2}}v}  \label{r10}
\end{equation}%
such that the line element (\ref{r8}) transforms into the following regular
(at the horizon) form%
\begin{equation}
ds^{2}=-\frac{\left( r+r_{h}\right) {}^{4}e^{\frac{2r_{h}\left(
3r+2r_{h}\right) }{\left( r+r_{h}\right) ^{2}}}}{R_{0}^{2}r^{2}}dUdV+R\left(
r\right) {}^{2}d\Omega ^{2}.  \label{r11}
\end{equation}%
The line element (\ref{r11}) is regular everywhere except at the origin
i.e., $r=0,$ and is valid for the entire region of $r>0$. From (\ref{r9})
and (\ref{r10}), one finds%
\begin{equation}
UV=-\left( \frac{R_{0}^{2}}{4r_{h}}\right) ^{2}e^{\frac{8r_{h}}{R_{0}^{2}}%
r_{\ast }}=-2\left( \frac{R_{0}^{2}}{4r_{h}}\right) ^{2}\frac{r-r_{h}}{%
r+r_{h}}e^{-\frac{2r_{h}\left( 3r+2r_{h}\right) }{\left( r+r_{h}\right) ^{2}}%
},  \label{r12}
\end{equation}%
and%
\begin{equation*}
\frac{V}{U}=e^{\frac{8r_{h}}{R_{0}^{2}}\left( \frac{v+w}{2}\right) }=e^{%
\frac{8r_{h}}{R_{0}^{2}}t},
\end{equation*}%
which help us to understand the nature of spacetime not only outside the
black hole but also inside. Eq. (\ref{r7}) implies that at the horizon where 
$r=r_{h},$ $r_{\ast }\rightarrow -\infty $ and $UV=0$. On the other hand
when $r\rightarrow \infty $, $r_{\ast }=0$ which implies $UV=-\left( \frac{%
R_{0}^{2}}{4r_{h}}\right) ^{2}<0$ which is a constant. More interestingly,
with $r\rightarrow 0$, $UV=2\left( \frac{R_{0}^{2}}{4er_{h}}\right) ^{2}>0$
which is also a constant implying the singularity is spacelike. Based on the
above calculations in Fig. \ref{F4}, we plot the Penrose diagram of the
black hole spacetime (\ref{I2}). It shows that the topology of the
asymptotic regions i.e., $I$ and $II$ is $AdS_{2}\times S^{2}$, however, the
radius of the round 2-sphere at a point of radius $r$ is not equal to $r$
but $R\left( r\right) ,$ given in (\ref{I4}). Moreover, the topology of the
interior region i.e., $III$ and $IV$ which are the future black hole and the
past black hole (or white hole) is $%
\mathbb{R}
\times 
\mathbb{R}
\times S^{2}$ such that the radius of $S^{2}$ is again $R\left( r\right) $.
We also noted in Fig. \ref{F4} the boundaries in terms of the transformed
coordinates in (\ref{r2}). It is clear that the black hole is bounded by a
timelike hypersurface located at $R=R_{0}.$ In this regard, the radius of
the 2-sphere corresponding to each point of radius $R$ is exactly $R.$

\section{Trajectories of particles}

\begin{figure}[tbp]
\includegraphics[scale=0.7]{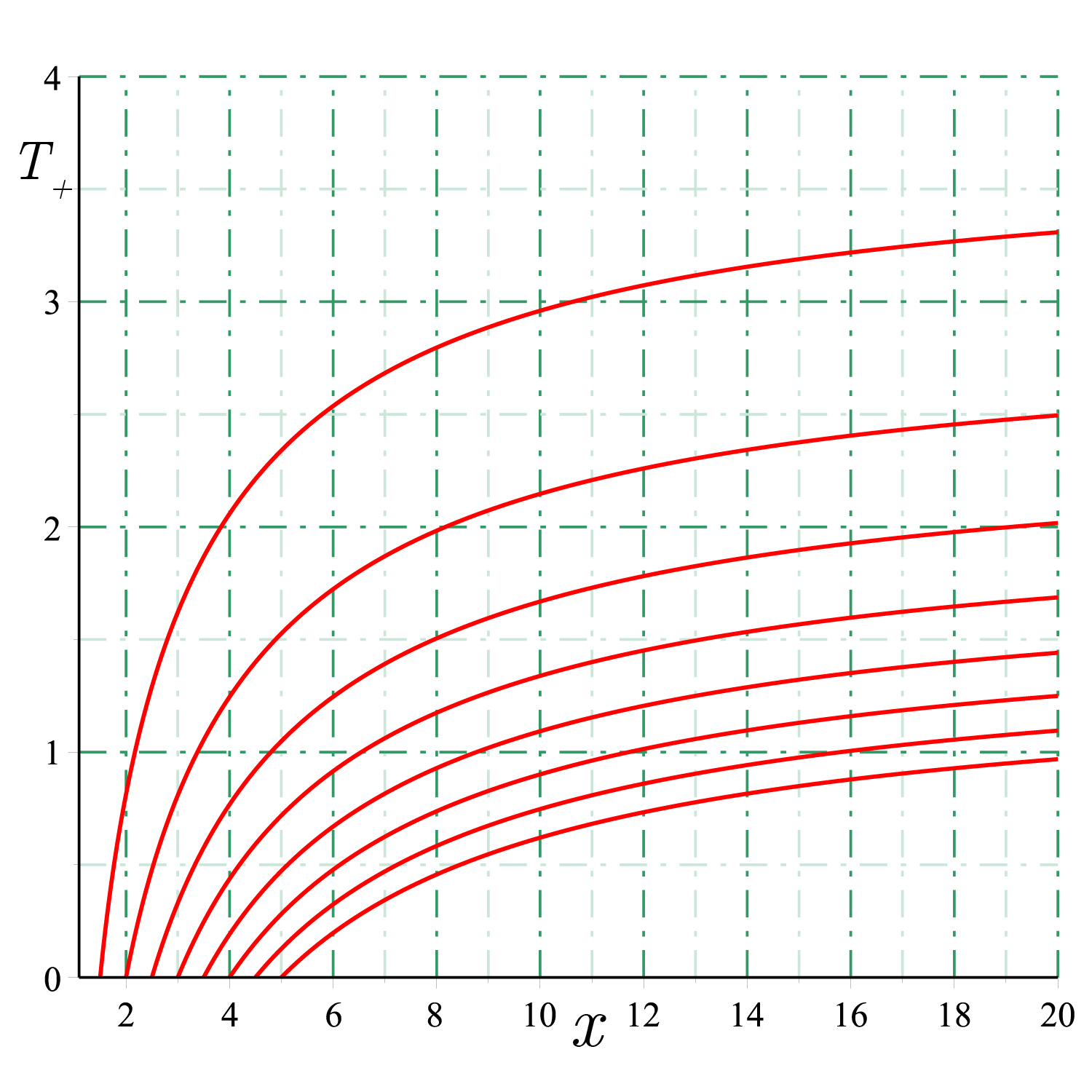}
\caption{The plots of $T_{+}$ in terms of $x$, from Eq. (\protect\ref{H1})
for $x_{0}=1.5,2.0,...,5$ with equal steps. }
\label{F5}
\end{figure}
\begin{figure}[tbp]
\includegraphics[scale=0.7]{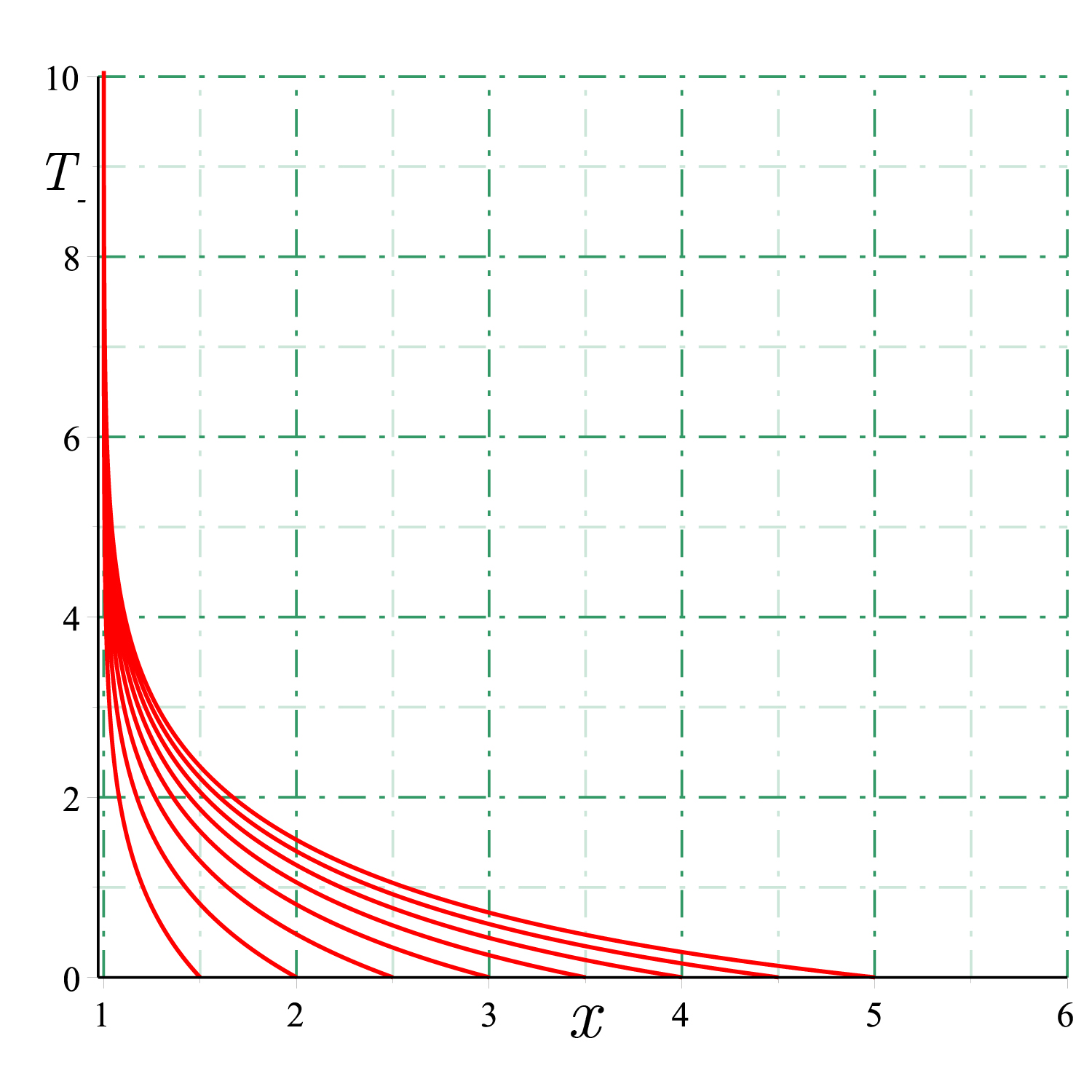}
\caption{The plots of $T_{-}$ in terms of $x$, from Eq. (\protect\ref{H1})
for $x_{0}=1.5,2.0,...,5$ with equal steps. }
\label{F6}
\end{figure}
\begin{figure}[tbp]
\includegraphics[scale=0.7]{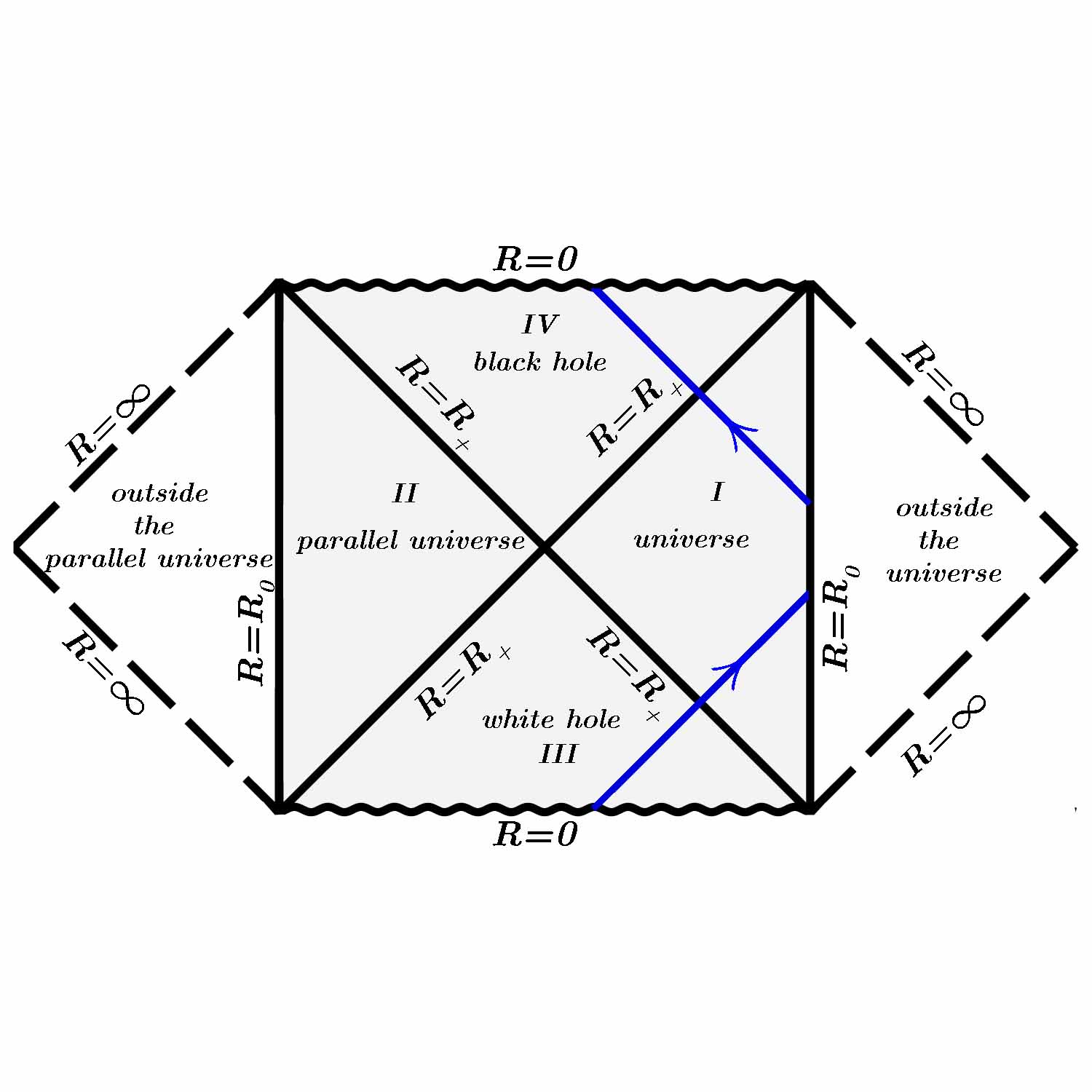}
\caption{The Penrose diagram of the black hole in the transformed
coordinates ($t,R,\protect\theta ,\protect\varphi $). The universe which is
supported by physical fields is shaded. The light beam can not reach outside
regions which are not physical.}
\label{F7}
\end{figure}

In this section, we study the geodesics of particles in the spacetime of the
black hole introduced in Ref. \cite{25}. Precisely, we investigate the
radial and circular time-like and null geodesics of test particles. The
static spherically symmetric black hole found in Ref. \cite{25} is described
by the line element (\ref{I2}) where the two metric functions are given by (%
\ref{I2}) and (\ref{I3}) in which $r\in \left[ 0,\infty \right) ,$ $R_{0}$
is a real constant and $r_{h}$ is the radius of the event horizon. The
Lagrangian of a massive particle with a unit mass is described by $\mathcal{L%
}=\frac{1}{2}g_{\mu \nu }\dot{x}^{\mu }\dot{x}^{\nu }$ which explicitly
reads 
\begin{equation}
\mathcal{L}=\frac{1}{2}\left( -f\left( r\right) \dot{t}^{2}+\frac{r^{2}}{%
f\left( r\right) }+R\left( r\right) {}^{2}\left( \dot{\theta}^{2}+\sin
^{2}\theta \dot{\phi}^{2}\right) \right)  \label{4}
\end{equation}%
in which a dot indicates a derivative with respect to an affine parameter
(here $\lambda $) for massless particles and the proper time for massive
particles. The Lagrangian (\ref{4}) is independent of time $t$ and azimuthal
angle $\phi $ which results in two conserved quantities: 1) the energy $E=-%
\frac{\partial \mathcal{L}}{\partial \dot{t}}$ and 2) the angular momentum
in $\phi $ direction $\ell =\frac{\partial \mathcal{L}}{\partial \dot{\phi}}$%
. Therefore, one writes

\begin{equation}
E=\,f\left( r\right) \frac{dt}{d\lambda }  \label{5}
\end{equation}%
and

\begin{equation}
\ell =\,R\left( r\right) ^{2}\sin ^{2}\theta \frac{d\phi }{d\lambda }
\label{6}
\end{equation}%
which are both conserved. Since we are interested in the geodesics on the
equatorial plane where $\theta \,=\,\frac{\pi }{2}$, the angular momentum
reduces to 
\begin{equation}
\ell =R\left( r\right) {}^{2}\frac{d\phi }{d\lambda }.  \label{7}
\end{equation}%
Furthermore, considering the definition of the four-velocity $U^{\mu }=\frac{%
dx^{\mu }}{d\lambda }$ that satisfies $U^{\mu }U_{\mu }=-\epsilon $ in which 
$\epsilon =+1,-1,$ and $0$ for timelike, spacelike, and null particles, we
explicitly obtain the condition

\begin{equation}
g_{\mu \nu }\frac{dx^{\mu }}{d\lambda }\frac{dx^{\nu }}{d\lambda }=-\epsilon
.  \label{8}
\end{equation}%
Upon combining Eqs. (\ref{5}), (\ref{6}) and (\ref{8}), after setting $%
\theta \,=\,\frac{\pi }{2},$ we determine

\begin{equation}
\left( \frac{dr}{d\lambda }\right) ^{2}=E^{2}-f\left( r\right) \left(
\epsilon +\frac{\ell ^{2}}{R(r)^{2}}\right)  \label{9}
\end{equation}%
which is the geodesic equation for the radial coordinate $r.$ In addition,
by introducing an effective potential in the form $V_{eff}^{2}\left(
r\right) \,=\,\frac{1}{2}f\left( r\right) \left( \epsilon +\frac{\ell ^{2}}{%
R(r)^{2}}\right) $ and an effective energy $\varepsilon _{eff}^{2}\,=\,\frac{%
1}{2}E^{2}$, the radial geodesics equation is given in a more familiar form
of an equation of motion for a test particle with unit mass, i.e.,

\begin{equation}
\frac{1}{2}\left( \frac{du}{d\lambda }\right) ^{2}+V_{eff}^{2}\left(
r\right) =\mathcal{E}_{eff}^{2}.  \label{10}
\end{equation}%
The exact form of the effective potential by replacing $f(r)$ from Eq. (\ref%
{I3}) is expressed by

\begin{equation}
V_{eff}^{2}\left( r\right) =\frac{1}{2}\frac{\left( r+r_{h}\right)
{}^{2}\left( r^{2}-r_{h}^{2}\right) }{R_{0}^{2}r^{2}}\left( \epsilon +\frac{%
\ell ^{2}\left( r+r_{h}\right) ^{2}}{R_{0}^{2}r^{2}}\right) .  \label{11}
\end{equation}%
Let us note that, from Eq. (\ref{10}) to have $r$ a real coordinate, the
condition $\mathcal{E}_{eff}^{2}\geq V_{eff}^{2}\left( r\right) $ should
hold. Moreover, by applying the chain rule $\frac{dr}{d\lambda }\,=\,\frac{dr%
}{d\phi }\frac{d\phi }{d\lambda }$, one eliminates the affine parameter from
the main geodesic equation to get

\begin{equation}
\left( \frac{dr}{d\phi }\right) ^{2}=\frac{2R\left( r\right) {}^{4}}{\ell
^{2}}\left( \mathcal{E}_{eff}^{2}-V_{eff}^{2}\left( r\right) \right) .
\label{12}
\end{equation}%
The latter equation explicitly reads

\begin{equation}
\left( \frac{dr}{d\phi }\right) ^{2}=\frac{E^{2}}{\ell ^{2}}R\left( r\right)
{}^{4}-f\left( r\right) \left( \epsilon \frac{R(r)^{4}}{\ell ^{2}}%
+R(r)^{2}\right) \,:=\,G\left( r\right)  \label{13}
\end{equation}%
in which the right-hand side has to satisfy $G\left( r\right) \geq 0$. In
what follows we classify the geodesics in different cases.

\subsection{Radial Motion}

\begin{figure}[tbp]
\includegraphics[scale=0.7]{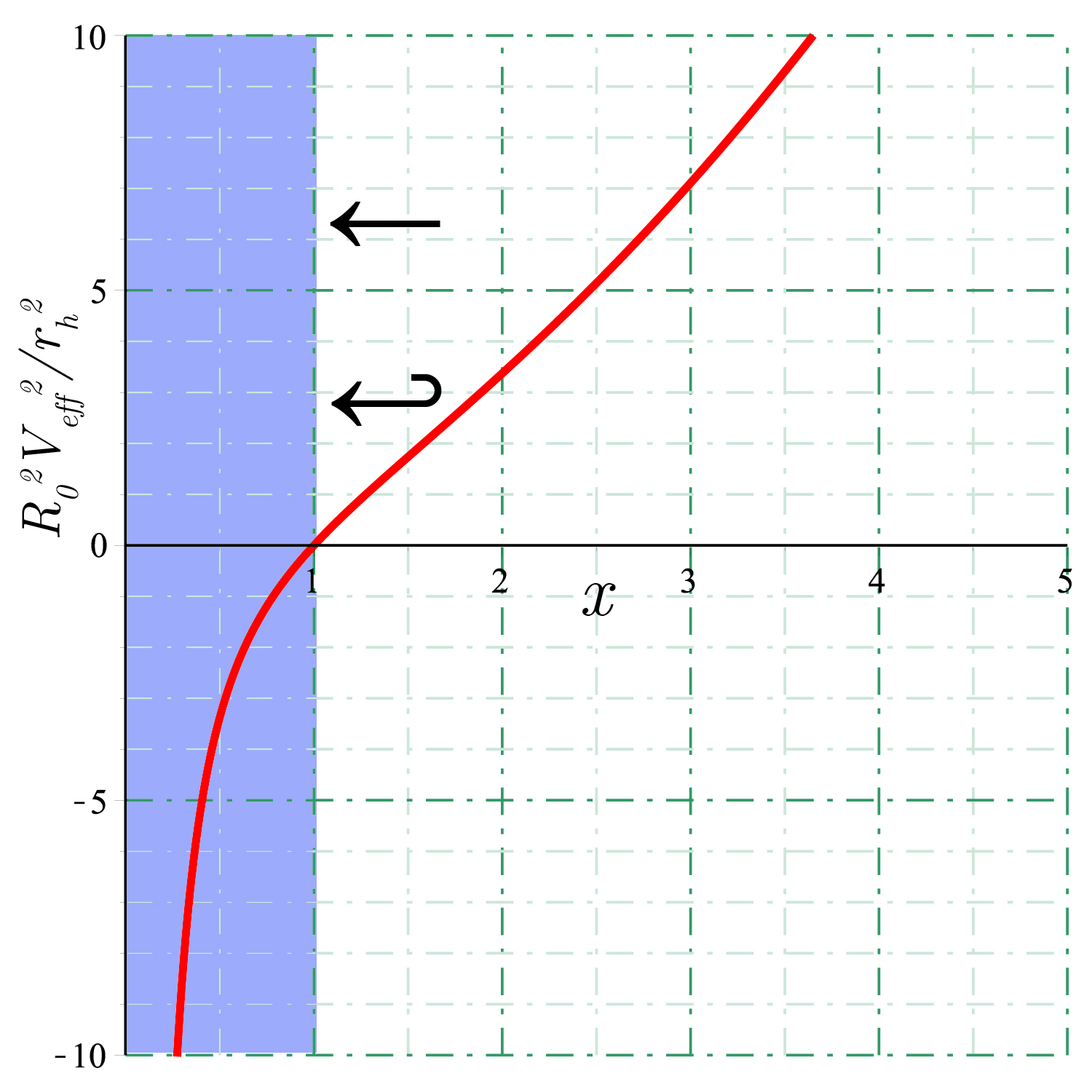}
\caption{The generic plot of $\frac{R_{0}^{2}}{u_{h}}V_{eff}^{2}\left(
r\right) $ in terms of $x=\frac{r}{r_{h}}.$ The shaded region implies inside
the black hole and the horizon is at $x=1.$ The massive particle in its
radial motion has no chance to escape to the edge of spacetime. This means
that the massive timelike particle either directly collapses to the
singularity of the spacetime or after it bounces from the potential barrier,
as shown in the figure. }
\label{F8}
\end{figure}

For a radial motion the angular momentum has to be zero i.e., $\ell \,=\,0$,
upon which (\ref{6}) yields $\phi \,=\,const.$ such that the particle will
move radially. Hence, Eq. (\ref{9}) reduces to 
\begin{equation}
\left( \frac{dr}{d\lambda }\right) ^{2}=E^{2}-\epsilon f\left( r\right)
\label{14}
\end{equation}%
In the following subsections, we shall investigate null and time-like
geodesics for radial motion separately.

\subsubsection{Null geodesics $\protect\epsilon\,=\,0$}

Considering $\epsilon \,=\,0$ in (\ref{14}) for the null geodesics which
describes the motion of a massless particle (photon), we simply find

\begin{equation}
\left( \frac{dr}{d\lambda }\right) ^{2}=E^{2}  \label{15}
\end{equation}%
which simply shows that the radial outgoing null geodesics are
future-complete. On the other hand combining (\ref{5}) and (\ref{15}), we
get,

\begin{equation}
\frac{dr}{dt}=\pm f\left( r\right) \,=\,\pm \frac{\left( r+r_{h}\right)
{}^{2}\left( r^{2}-r_{h}^{2}\right) }{R_{0}^{2}r^{2}}.  \label{16}
\end{equation}%
Eq. (\ref{16}) is integrable and explicitly yields the following 
\begin{equation}
\frac{R_{0}^{2}}{4}\left( \frac{1}{2r_{h}}\ln \left[ \frac{\left(
r-r_{h}\right) \left( r_{0}+r_{h}\right) }{\left( r+r_{h}\right) \left(
r_{0}-r_{h}\right) }\right] -\frac{3r+2r_{h}}{\left( r+r_{h}\right) {}^{2}}+%
\frac{3r_{0}+2r_{h}}{\left( r_{0}+r_{h}\right) {}^{2}}\right) =\pm \left(
t-t_{0}\right)  \label{17}
\end{equation}%
where $t_{0}$ is the initial time, $t$ is the time measured by the distant
observer and $r_{0}$ is the initial position of the massless particle
(photon). Introducing $r=r_{h}x$, $r_{0}=r_{h}x_{0}$ and $T=\frac{8r_{h}}{%
R_{0}^{2}}\left( t-t_{0}\right) $ we obtain%
\begin{equation}
T_{\pm }=\pm \left( \ln \left[ \frac{\left( x-1\right) \left( x_{0}+1\right) 
}{\left( x+1\right) \left( x_{0}-1\right) }\right] -2\frac{3x+2}{\left(
x+1\right) {}^{2}}+2\frac{3x_{0}+2}{\left( x_{0}+1\right) {}^{2}}\right) .
\label{H1}
\end{equation}%
We note that $\pm $ refers to the outgoing or ingoing light rays. In Fig. %
\ref{F5} we plot $T_{+}$ in terms of $x$ for various values of $x_{0}.$ From
this figure, we observe that on the equilateral plane, the photon moves away
from the horizon toward the boundary of the spacetime where $r\rightarrow
\infty .$ The time needed for the photon to reach the boundary of the
spacetime is found to be%
\begin{equation}
T_{+\infty }=2\frac{3x_{0}+2}{\left( x_{0}+1\right) {}^{2}}+\ln \frac{x_{0}+1%
}{x_{0}-1}  \label{H2}
\end{equation}%
which is finite. Furthermore, in Fig. \ref{F6} we plot $T$ in terms of $x$
for various values of $x_{0}$. We see that with initial velocity toward the
horizon, the photon to reach the horizon at an infinite time measured by a
distant observer.

\subsubsection{Null geodesics in the transformed coordinates}

In this part, we would like to present the radial motion of null particles
namely photons in the vicinity of the black hole described by the
transformed coordinate given in Eqs. (\ref{r2})-(\ref{r4}). Similar
calculations reveal that the transformed version of the radial equation for
a null particle i.e., Eq. (\ref{15}) becomes%
\begin{equation}
\left( \frac{dR}{d\lambda }\right) ^{2}=\frac{h\left( R\right) }{f\left(
R\right) }E^{2}.  \label{N1}
\end{equation}%
This equation explicitly reads as%
\begin{equation}
\frac{dR}{d\lambda }=\pm \frac{\left\vert E\right\vert }{u_{h}R_{0}}\left(
R_{0}-R\right) ^{2}  \label{N2}
\end{equation}%
in which $\pm $ stand for the outgoing ($+$) and ingoing ($-$) null beam.
Assuming the initial position of the null beam to be $R=R_{i}$ at $\lambda
=\lambda _{i},$ the latter equation admits exact solutions given by%
\begin{equation}
R_{\pm }=R_{0}-\frac{1}{\frac{1}{R_{0}-R_{i}}\pm \frac{\left\vert
E\right\vert }{u_{h}R_{0}}\left( \lambda -\lambda _{i}\right) }.  \label{N3}
\end{equation}%
One can easily check that $\lim_{\lambda \rightarrow \lambda _{i}}R_{\pm
}=R_{i}$, while%
\begin{equation}
\lim_{\lambda \rightarrow \infty }R_{+}=R_{0},  \label{N4}
\end{equation}%
which implies that null particles cannot pass the boundary at $R=R_{0}.$ On
the other hand, (\ref{N3}) yields the interval of $\lambda $ which is needed
for the ingoing beam reaches to the event horizon given by%
\begin{equation}
\triangle \lambda =\frac{u_{h}R_{0}}{\left\vert E\right\vert }\left( \frac{1%
}{R_{0}-R_{i}}-\frac{2}{R_{0}}\right) .  \label{N5}
\end{equation}%
This shows that the maximum $\triangle \lambda \rightarrow \infty $ when $%
R_{i}\rightarrow R_{0}$ when the light beam is directed toward the black
hole from the edge of the universe. For any other initial position, $%
\triangle \lambda $ is a finite positive number. In Fig. \ref{F7}, we plot
the Penrose diagram of the black hole in the transformed coordinates with
the line element (\ref{r2}). As expected the Penrose diagram is the same as
Fig. \ref{F4} and the only difference is the boundary of the universe.
Furthermore, in Fig. \ref{F7} we have shown the excluded regions i.e., $%
R>R_{0}$ which corresponds to the phantom scalar field and the light ray can
not reach there. The blue rays stand for the null geodesics.

\subsubsection{Time-like geodesics $\protect\epsilon \,=\,1$}

\begin{figure}[tbp]
\includegraphics[scale=0.7]{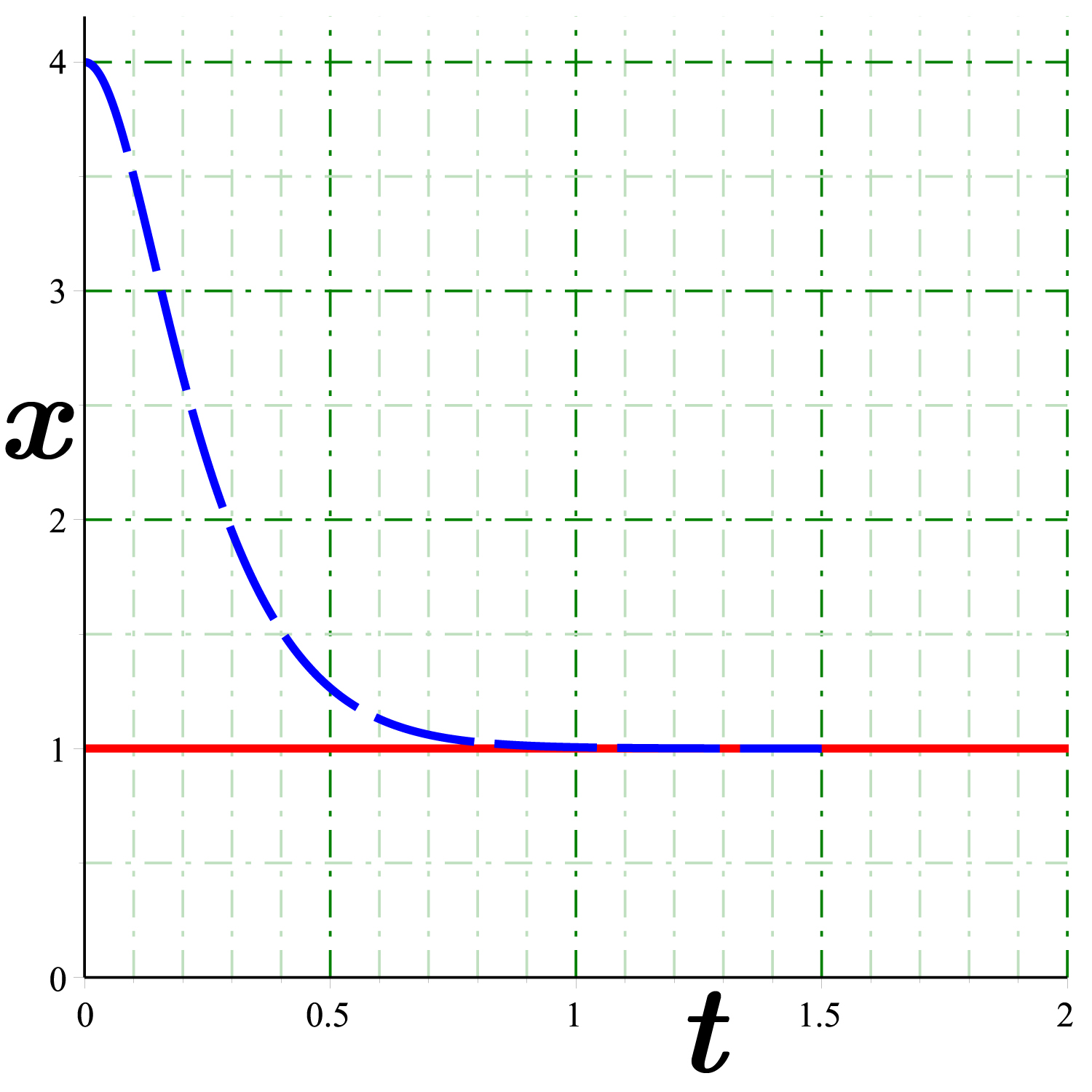}
\caption{The radial motion of a massive particle (dashed-blue curve) with $%
\tilde{R}_{0}=1,\left( \frac{dx}{dt}\right) _{t=0}=0$ and $\tilde{E}^{2}=%
\frac{375}{16}$. The horizon is located at $x=1$ (red solid line). This is
the numeric plot of the radial geodesic equation (\protect\ref{R2})
constraint by (\protect\ref{R1}).}
\label{F9}
\end{figure}

\begin{figure}[tbp]
\includegraphics[scale=0.7]{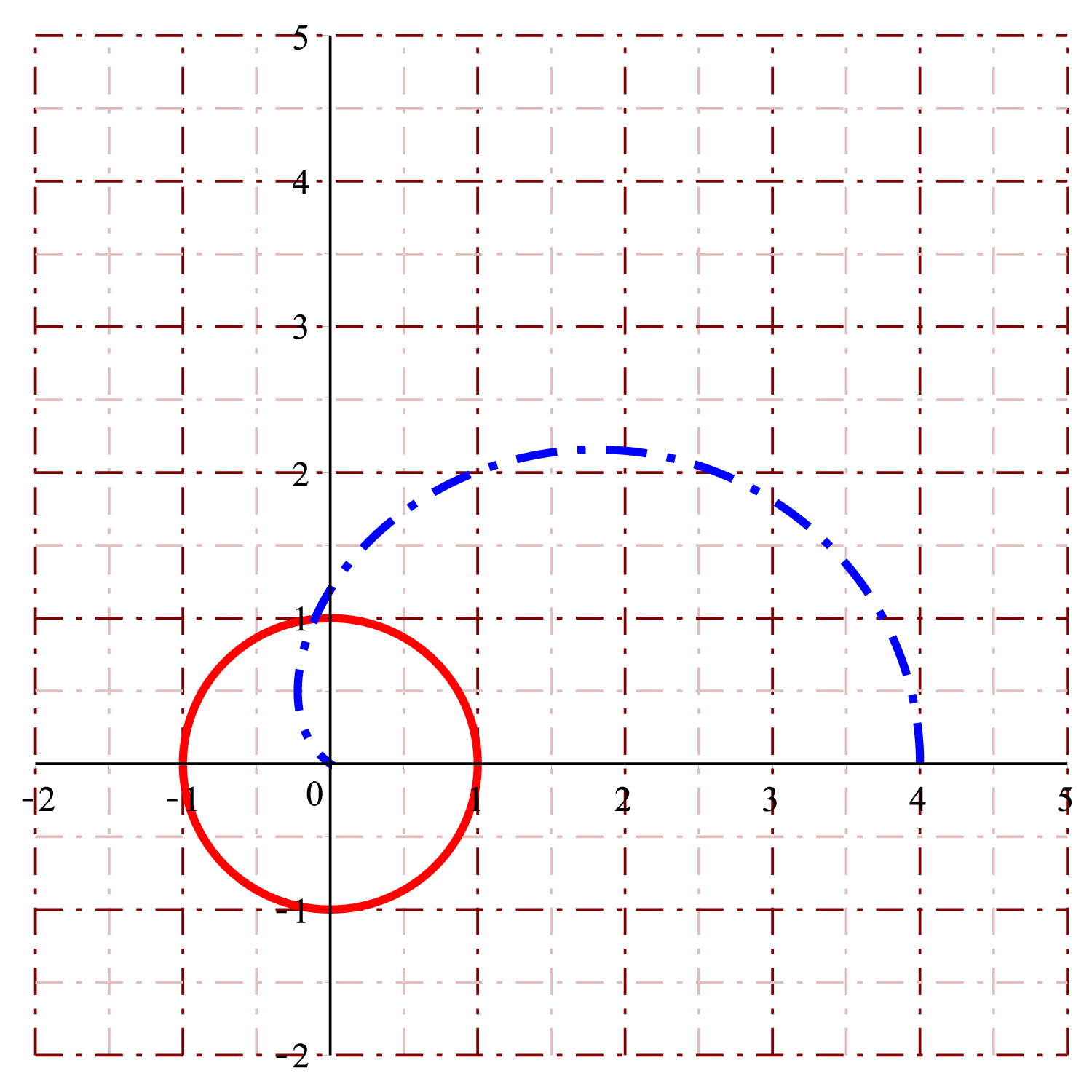}
\caption{Trajectory of a photon (dashed-blue curve) with $\tilde{\ell}%
^{2}=2,\left( \frac{dx}{d\protect\varphi }\right) _{\protect\varphi =0}=0$
and $\tilde{E}^{2}=\frac{9375}{128}$. The horizon is located at $x=1$ (red
circle).}
\label{F10}
\end{figure}
\begin{figure}[tbp]
\includegraphics[scale=0.7]{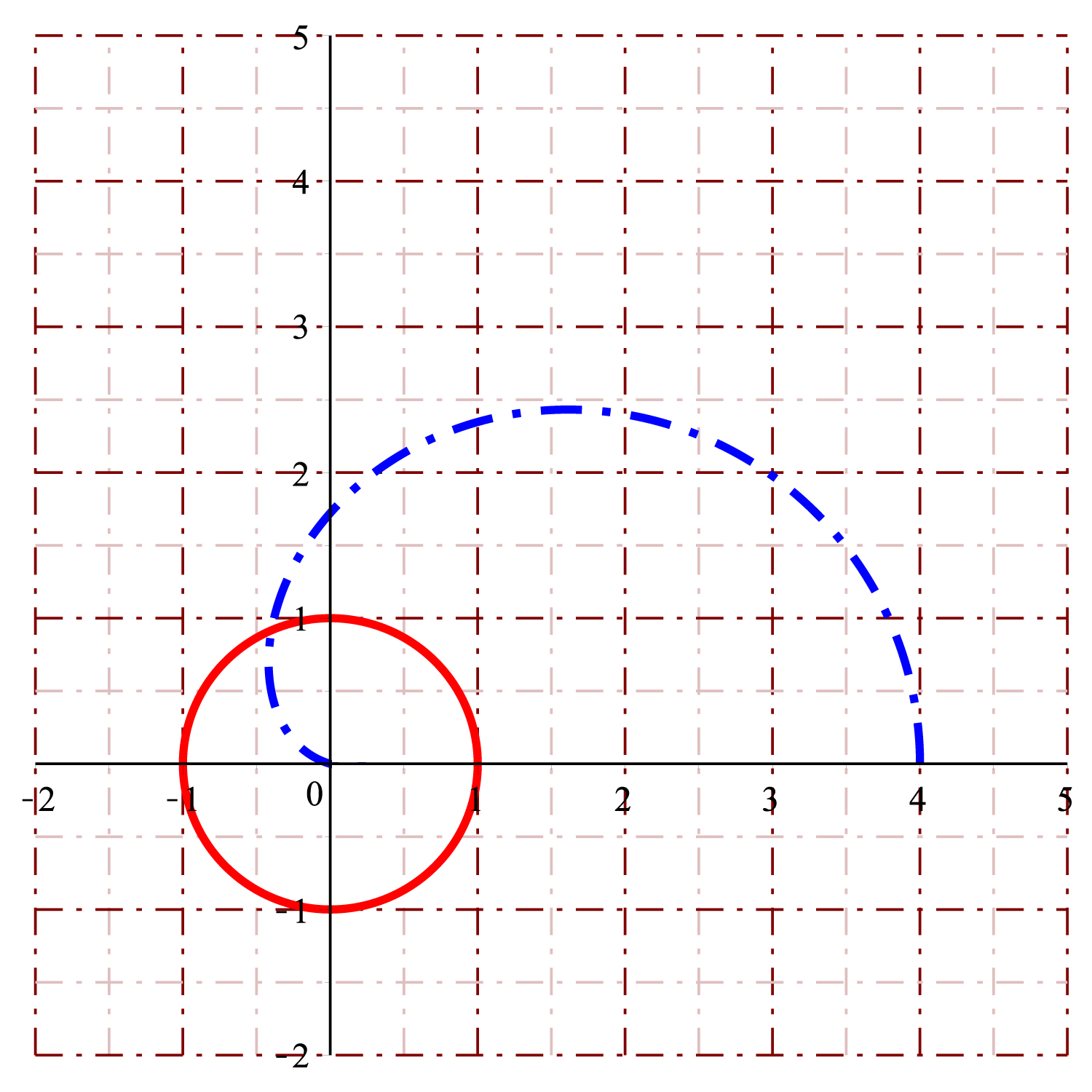}
\caption{Trajectory of a timelike particle (dashed-blue curve) with $\tilde{%
\ell}^{2}=2,\left( \frac{dx}{d\protect\varphi }\right) _{\protect\varphi %
=0}=0$ and $\tilde{E}^{2}=\frac{12375}{128}$. The horizon is located at $x=1$
(red circle).}
\label{F11}
\end{figure}

Time-like geodesics refers to the motion of a massive particle where $%
\epsilon \,=\,1$ upon which (\ref{14}) becomes

\begin{equation}
\left( \frac{dr}{d\lambda }\right) ^{2}=\,E^{2}-\frac{\left( r+r_{h}\right)
{}^{2}\left( r^{2}-r_{h}^{2}\right) }{R_{0}^{2}r^{2}}.  \label{18}
\end{equation}%
Derivative of (\ref{18}) with respect to the affine parameter $\lambda $
implies

\begin{equation}
\frac{d^{2}r}{d\tau ^{2}}=-\frac{1}{R_{0}^{2}r^{3}}\left( r+r_{h}\right)
\left( r^{3}+r_{h}^{3}\right) ,  \label{19}
\end{equation}%
in which we set $\lambda =\tau $ where $\tau $ is the proper time.
Obviously, the radial force per unit mass is attractive and toward the
horizon of the black hole. We assume that the particle is initially at rest
located at $r=r_{0}$ such that (\ref{18}) yields

\begin{equation}
E^{2}=\frac{\left( r_{0}+r_{h}\right) {}^{2}\left(
r_{0}^{2}-r_{h}^{2}\right) }{R_{0}^{2}r_{0}^{2}}  \label{21}
\end{equation}%
upon which (\ref{18}) becomes

\begin{equation}
\left( \frac{dr}{d\tau }\right) ^{2}=\frac{1}{R_{0}^{2}}\left( \frac{\left(
r_{0}+r_{h}\right) {}^{2}\left( r_{0}^{2}-r_{h}^{2}\right) }{r_{0}^{2}}-%
\frac{\left( r+r_{h}\right) {}^{2}\left( r^{2}-r_{h}^{2}\right) }{r^{2}}%
\right) ,  \label{22}
\end{equation}%
and with $\ell \,=\,0$ and $\epsilon \,=\,1$, the effective potential becomes

\begin{equation}
V_{eff}^{2}\left( r\right) =\frac{1}{2}\frac{\left( r+r_{h}\right)
{}^{2}\left( r^{2}-r_{h}^{2}\right) }{R_{0}^{2}r^{2}}.  \label{23}
\end{equation}%
In Fig. \ref{F8} we plotted $\frac{R_{0}^{2}}{r_{h}}V_{eff}^{2}\left(
r\right) $ in terms of $x=\frac{r}{r_{h}}$ which shows that the potential is
an increasing function implying an attractive force toward the singularity.
Therefore no matter what is the energy of the particle, its fate is a
collapse into the singularity. Finally, using Eqs. (\ref{18}) and (\ref{5})
we find the radial equation of motion of a massive particle in terms of the
observer time i.e.,

\begin{equation}
\left( \frac{dr}{dt}\right) ^{2}=f\left( r\right) {}^{2}-\frac{f\left(
r\right) {}^{3}}{E^{2}}  \label{24}
\end{equation}%
and explicitly

\begin{equation}
\left( \frac{dr}{dt}\right) ^{2}=\frac{\left( r+r_{h}\right) {}^{4}\left(
r^{2}-r_{h}^{2}\right) {}^{2}}{R_{0}^{4}r^{4}}\left( 1-\frac{\left(
r+r_{h}\right) {}^{2}\left( r^{2}-r_{h}^{2}\right) }{E^{2}R_{0}^{2}r^{2}}%
\right) .  \label{25}
\end{equation}%
Considering the timelike particle is initially at rest where $r=r_{0}$, one
writes

\begin{equation}
0=\frac{\left( r_{0}+r_{h}\right) {}^{4}\left( r_{0}^{2}-r_{h}^{2}\right)
{}^{2}}{R_{0}^{4}r_{0}^{4}}\left( 1-\frac{\left( r_{0}+r_{h}\right)
{}^{2}\left( r_{0}^{2}-r_{h}^{2}\right) }{E^{2}R_{0}^{2}r_{0}^{2}}\right)
\label{27}
\end{equation}%
which yields the conserved energy of the particle in terms of its initial
position given in Eq. (\ref{21}). Introducing $r\left( t\right) =x\left(
t\right) r_{h}$, $R_{0}=\tilde{R}_{0}\sqrt{r_{h}}$ and $E=\tilde{E}\sqrt{%
r_{h}}$ we obtain the following equation%
\begin{equation}
\left( \frac{dx}{dt}\right) ^{2}=\frac{\left( x^{2}-1\right) ^{2}\left(
1+x\right) ^{4}}{\tilde{R}_{0}^{4}x^{4}}-\frac{\left( x^{2}-1\right)
^{3}\left( 1+x\right) ^{6}}{\tilde{R}_{0}^{6}\tilde{E}^{2}x^{6}}  \label{R1}
\end{equation}%
and after differentiating with respect to $t$ the main differential equation
is obtained 
\begin{equation}
\frac{d^{2}x}{dt^{2}}=\frac{2\left( x^{2}-1\right) \left( 1+x\right)
^{4}\left( 1-x+x^{2}\right) }{\tilde{R}_{0}^{4}x^{5}}-\frac{3\left(
x^{2}-1\right) ^{2}\left( 1+x\right) ^{6}\left( 1-x+x^{2}\right) }{\tilde{R}%
_{0}^{6}\tilde{E}^{2}x^{7}}.  \label{R2}
\end{equation}%
The latter equation is highly nonlinear and cannot be solved analytically.
Therefore we solve it by applying the numerical method. In Fig. \ref{F9} we
plotted $x\left( t\right) $ in terms of $t$ where the parameters are set to
be $\tilde{R}_{0}=1$ and $\tilde{E}^{2}=\frac{375}{16}$ and the initial
conditions are given by $x_{0}=4$ and $\left( \frac{dx}{dt}\right) _{0}=0.$

\subsection{General equatorial motion}

Having known that a circular motion is not possible, in this section we
study the general motion of null/timelike particles on the equatorial plane.
The trajectory of such particles has to satisfy the following equations%
\begin{equation}
\left( \frac{dx}{d\varphi }\right) ^{2}=\epsilon \frac{\left( 1-x\right)
x^{2}}{\left( 1+x\right) \tilde{\ell}^{2}}+\frac{\tilde{E}^{2}x^{4}}{\tilde{%
\ell}^{2}\left( 1+x\right) ^{4}}+\left( 1-x^{2}\right)  \label{34}
\end{equation}%
and%
\begin{equation}
\frac{d^{2}x}{d\varphi ^{2}}=\epsilon \frac{\left( 1-x-x^{2}\right) x}{%
\left( 1+x\right) ^{2}\tilde{\ell}^{2}}+\frac{2\tilde{E}^{2}x^{3}}{\tilde{%
\ell}^{2}\left( 1+x\right) ^{5}}-x  \label{35}
\end{equation}%
in which we have introduced $x\left( \varphi \right) =r\left( \varphi
\right) /r_{h},$ $\tilde{\ell}^{2}=\ell ^{2}/R_{0}^{2},$ and $\tilde{E}%
^{2}=R_{0}^{2}E^{2}/r_{h}^{2}.$ The nonlinear structure of the geodesics
main equation (\ref{35}) prevents obtaining an exact solution, however, we
solved (\ref{35}) using a numerical method, and the results are presented in
Figs. \ref{F10} and \ref{F11} for the null ($\epsilon =0$) and the timelike (%
$\epsilon =1$) particles. It is observed that with nonzero angular momentum,
no matter what the initial conditions are, the particle eventually falls
into the black hole.

\section{Conclusion}

This study sheds light on the physical properties of the black hole of Ref. 
\cite{25} which is of a class of black holes formed in $S_{3}$ background
spatial topology. In the $r$ coordinate, one can see from the line element (%
\ref{I2}) and the radial null geodesic equation (\ref{15}) that the
spacetime is not compact. This can also be seen in the Penrose diagram of
the black hole solution in Figs. \ref{F4}. However, the radius of the $S^{2}$
sphere corresponding to each point of spacetime of radius $r$ is $R\left(
r\right) =R_{0}\frac{r}{r+r_{h}}$ that asymptotically goes to $R_{0}.$ With
a proper transformation we have shown that in the transformed line element
given in Eq. (\ref{K6}), the spatial part possesses an $S_{3}$ topology in
the vicinity of $\chi \simeq \frac{14835\pi }{65536}$. Our new analysis in
the transformed line element, particularly based on the radial null
geodesics equation (\ref{N2}), shows that the light beam can not pass the
timelike boundary of the black hole at $R=R_{0}.$ This is demonstrated in
the second Penrose diagram in Fig. \ref{F7}. In addition to that we studied
the geodesics of the null and timelike particles. In the first part of our
geodesics analysis, we studied the radial motion of a photon and a massive
particle on the equatorial plane. For a photon, we observed that depending
on the light beam's direction, it may go to the edge of the universe in a
finite time measured by a distant observer or it reaches the black hole in
an infinite time. On the other hand, a massive timelike particle collapses
to the singularity irrespective of the direction of its initial velocity.
One can see from Fig. \ref{F9} that there is no stable circular orbit
neither for a photon nor for a massive particle. We observed also that
massive particles as well as non-radially outgoing light beams all are
attracted by the black hole. Is this unusual or universal behavior of all
black holes formed in $S_{3}$ background spatial topology? This question can
not be answered here. A concrete answer needs further investigation on the
geodesics of other black holes in the same class. We leave this problem open.

\end{document}